\documentclass[aip, reprint, superscriptaddress]{revtex4-1}

\usepackage{todonotes}
\usepackage{bm}
\usepackage{graphicx}
\usepackage{amsmath}
\usepackage{mathbbol}
\usepackage{amssymb}
\usepackage[final]{listings}
\usepackage[normalem]{ulem}
\usepackage{color}
\usepackage{hyperref}

\definecolor{bg}{rgb}{0.9,0.9,0.9}

\begin{document}
\lstset{language=Python,
        backgroundcolor=\color{bg},
        basicstyle=\fontsize{9.5}{10}\ttfamily,
        aboveskip=2pt,
        frame=tlbr,
        framesep=2pt,
        framerule=0pt,
        keywordstyle=\color{blue},
        commentstyle=\itshape\color{cyan},
        stringstyle=\color{red},
        breaklines=true,
        columns=fullflexible,
        morekeywords={as},
}

\title{User interfaces for computational science: a domain
  specific language for OOMMF embedded in Python}

\author{Marijan Beg} \email{m.beg@soton.ac.uk}
\author{Ryan A. Pepper}
\author{Hans Fangohr} \email{h.fangohr@soton.ac.uk}
\affiliation{$^{1}$Faculty of Engineering and the Environment, University of
  Southampton, Southampton, SO17 1BJ, United Kingdom}

\begin{abstract}
  Computer simulations are used widely across the engineering and
  science disciplines, including in the research and development of
  magnetic devices using computational micromagnetics. In this work,
  we identify and review different approaches to configuring
  simulation runs: (i) the re-compilation of source code, (ii) the use
  of configuration files, (iii) the graphical user interface,
  and (iv) embedding the simulation specification in an existing
  programming language to express the
  computational problem. We identify the advantages and disadvantages
  of different approaches and discuss their implications on
  effectiveness and reproducibility of computational studies and
  results. Following on from this, we design and describe a domain specific
  language for micromagnetics that is embedded in the Python language,
  and allows users to define the micromagnetic simulations they want
  to carry out in a flexible way. We have implemented this
  micromagnetic simulation description language together with a
  computational backend that executes the simulation task using the
  Object Oriented MicroMagnetic Framework (OOMMF). We illustrate the
  use of this Python interface for OOMMF by solving the micromagnetic
  standard problem 4. All the code is publicly available and is open source.
\end{abstract}

\maketitle
\section{Introduction}
Computational Science is emerging as the third pillar of research and
development in academia and in industry across all science and
engineering disciplines. Computational studies complement experimental
and theoretical studies, and are at times the only feasible way to
address research challenges, effective industrial design and
engineering of various products and systems.

In the field of magnetism, micromagnetic simulations have become well
established and are often the only possible technique for the
exploration of different magnetic phenomena.~\cite{Najafi2009,
  Venkat2013, Baker2017} Their use becomes more widespread and
reliable as the micromagnetic models, simulation techniques, and the
processing power of computers advance.

Computational science brings its own challenges: results based on
computer simulation should be reproducible,~\cite{Monya2016, Peng2011} ideally
by the whole research community but at the very
least by the authors of the publication. This requires, amongst
other things, tracking of all input parameters for a simulation,
and all post-processing steps, often for very many simulation
runs.~\cite{Freire2008}

In this work, we provide an overview of different approaches to
configuring simulations in Sec.~\ref{sec:simulation-user-interfaces}:
code re-compilation, the use of configuration scripts, graphical user
interfaces, as well as the use of domain specific languages to specify
the computational problem. We discuss advantages and disadvantages
with particular focus on the reproducibility associated with these
approaches. In Sec.~\ref{sec:python-interface}, we use the most
flexible approach identified and describe the design of a
Python~\cite{vanRossum2001} interface for the widely used Object
Oriented MicroMagnetic Framework (OOMMF) simulation
tool.~\cite{Donahue1999} Finally, we illustrate the use of this
interface by solving the micromagnetic standard problem~\cite{McMichael} 4 in
Sec.~\ref{sec:example}, before closing with a summary.

\section{Simulation User Interfaces}
\label{sec:simulation-user-interfaces}

Micromagnetic simulations, as with simulations in many other fields, need
to simulate the behaviour of particular materials with their specific
material constants, under particular circumstances (for example with an applied
field) and specific geometries. We refer to this parameter set as
the \emph{simulation input parameters}. Despite changing input
parameters, the computational framework (here the micromagnetic model)
stays roughly the same for all possible parameter sets (although
sometimes extra terms need to be added).

The challenges for the researcher using simulations include:
(i) communicating the input parameters to the
simulation, and (ii) keeping track of the input parameters that were
used for particular simulation results (for publication,
and to be able to repeat the exact simulation in the future).
We describe 4 approaches that have emerged over time.

\paragraph*{Approach 1: Code re-compilation.} The first approach for
providing input parameters to a computer simulation is to hard-code
the simulation parameters in the source code of the simulation, and re-compile
the simulation tool source code for each set of input
parameters. While easy to implement, it makes it difficult to store
the input parameters in an efficient way (unless the full copy of
the source code is archived for every simulation run).

\paragraph*{Approach 2: Configuration files.}
The second approach is the writing of the input parameters to a configuration file,
which is then read by the simulation tool at run time. There are several
advantages of this over the re-compilation approach. First, multiple
simulations can be run using the same simulation executable by providing
multiple configuration files, each describing material
properties and geometry for one simulation. Secondly, if simulation
configuration files as well as the simulation software version are kept,
all simulations can be repeated at a later stage. This is important for
reproducibility of the results; a topic receiving increasing
(and well deserved) attention in computational science.~\cite{Peng2011, Monya2016}
A disadvantage of the configuration-file based simulation
configuration approach is that the developers may have
to define a syntax for the configuration files and implement a parser
for it. Depending on the complexity of the syntax, the number of
different operations the simulation
tool can be configured to perform is limited. Experience shows that
this syntax (and parser) is often extended as the simulation tool's
capabilities grow and demand more
flexibility. Magpar~\cite{Scholz2003a} is a micromagnetic simulation
tool using this approach. OOMMF~\cite{Donahue1999} also uses
configuration files, but is using an existing language and parser (Tcl).

\paragraph*{Approach 3: Graphical User Interface.}
Simulation tools for which input parameters are set through
graphical user interfaces are often perceived as particularly
user friendly. If the completed configuration can be saved to a
file and reloaded later, this is similar to the
configuration file based approaches. Where the configuration cannot
be saved, it is difficult to reliably record all configuration
options as this would have to be done manually, and tedious
to re-execute a particular simulation as all input parameters need to be
entered again manually.

\paragraph*{Approach 4: Domain Specific Language (DSL) embedded in general purpose language.}
It is also possible to embed all input parameter definition and
high-level simulation commands in an executable file which fully defines the
simulation, using the syntax of an existing programming language.
The simulation is carried out by executing this file with the
appropriate interpreter or compiler. The one file contains all
the information that needs to be preserved for reproducibility.
Examples of embedding micromagnetic simulation tools into existing
programming language include Nmag,~\cite{Fischbacher2007}
Micromagnum,~\cite{Micromagnum} Magnum.fe,~\cite{Abert2013a} and
Fidimag.~\cite{Wang2016} An important advantages of this method over the
configuration file based approach is the increased flexibility:
as the simulation script contains a sequence of 'normal' commands in a
given programming language, these can be used and combined as required
to, for example, create complicated spatial field distributions, fetch
data from a file or a connected experimental kit, and do parameter
sweeps automatically \emph{within the same file}. It is possible to
carry out postprocessing within the same simulation file; thus keeping
input data, simulation process and extraction of results closely
together (supporting tracking of the provenance~\cite{Freire2008}).
If the chosen programming language is an interpreted one (such as
Python~\cite{vanRossum2001}), this embedded approach may also allow
\emph{interactive execution} of simulations (sometimes referred to as
``computational steering''), interactive analysis, and visualisation
(for instance, in Jupyter notebooks~\cite{Perez2007}).  Eventually, by
using an existing language, both code developers and users can benefit
from using existing and well tested modules for the selected
programming language. A disadvantage of this
approach is that it requires more planning on the coding side to
provide the described framework that can be used flexibly.

\paragraph*{Discussion.}
It is known from software engineering that there are significant
advantages of separating (configuration) metadata from (simulation)
software code to obtain programs that are more flexible and robust,
easier to maintain and test, and more versatile in their
use.\cite{Hunt1999} In our context, this separation of metadata
from the simulation code is given in the configuration-file approach
and in the embedded language approach; the latter providing more
flexibility in driving the simulation and integrating other steps of
the computational work flow.

\section{Python interface for OOMMF}
\label{sec:python-interface}

\paragraph*{Introduction.}
In what follows, we describe a design for a Domain Specific
Language (DSL) for micromagnetic simulations that is embedded in the
Python language. This allows scientists from the domain of
micromagnetics to express their simulation requirements using this
language, and the language is valid Python. We show an example in Sec.~\ref{sec:example}.
We have implemented a tool~\cite{oommfc}
that can understand this domain specific language and carry out the
required simulation using
OOMMF. This allows to express the micromagnetic simulation requirements in Python and
to postprocess and analyse data with great flexibility in Python, while the
computation is done by OOMMF.

\paragraph*{Choice of Python to embed the micromagnetic specification
  language.}
Here we explain the choice of Python as the implementation language for the
interface. Python~\cite{vanRossum2001} has been gaining popularity
in computational science since soon after its inception in
1991. Python has been identified as a language that is easy to
learn~\cite{Fangohr2004} by computational scientists and engineers.
The value of Python for computational science is its flexibility and
readability -- both attributes that reduce the time required to
express algorithms (including post-processing and plotting
instructions) and debugging them. While Python code -- if used
natively and naively -- can be orders of magnitude slower than C or
Fortran code, it is possible to develop and drive High Performance
Computing projects in Python.\cite{Gorelick2014} Finally, Python has
a rich variety of high quality and well tested modules providing
algorithms for performing common operations in computational science
and engineering, such as SciPy,~\cite{scipy} NumPy,~\cite{Walt2011}
Pandas,~\cite{McKinney2011} matplotlib,~\cite{Hunter2007} that we use
in this work.

\paragraph*{The choice of OOMMF as the computational backend.}
The Object Oriented MicroMagnetic Framework (OOMMF)~\cite{Donahue1999}
developed at the Information Technology Laboratory at the National
Institute of Standards and Technology by Michael J. Donahue and Donald G. Porter
is widely used. Its discretisation scheme is based on the finite
difference method. The computational core is written in C++, and
combined with Tcl/Tk for high level interfaces and Graphical User
Interfaces; combining tools in a clever way and using state-of-the art
technology at the time of OOMMF's inception. OOMMF uses predefined
simulation modes (for instance, hysteresis or dynamics), and does not
allow the user to carry out micromagnetic operations in an arbitrary
order. OOMMF's configuration files use the Tcl syntax, and thus allow
a convenient way to compute spatially distributed fields from
equations within that configuration file. However, OOMMF is not
embedded into an existing programming language which means, for
example, that no single configuration file can carry out a parameter
sweep, or host multiple simulation objects, and any post-processing
and visualisation must be carried out separately. We believe that
OOMMF's computational capabilities are highly valued by the community,
and it is likely to be the most widely used micromagnetic simulation code.

\paragraph*{Low level OOMMF interface implementation choice.}
Here, we describe how the implementation communicates with OOMMF. This is
transparent to the scientists using the specification language and
targeting a more technical audience of simulation tool developers.
The core computational routines of OOMMF are implemented in C++ but
important higher level functionality is written in Tcl, including
platform specific installation. We have researched and considered a
variety of technical solutions for interfacing Tcl and C++ code with
Python, including handcoding an interface, using
SWIG,~\cite{Beazley1996} Boost,~\cite{Abrahams2003} and
Cython.~\cite{Behnel2011} Eventually, we have opted for communication
with OOMMF via \texttt{mif} configuration and output data files. The
main advantage of this approach is the robustness. For example, we do
not need to adopt this interface code depending on the platform
(operating system and compiler) on which the OOMMF code was
compiled. We can deploy the Python interface code on a system where
there is no C++ compiler (but only the OOMMF executable). Furthermore,
the coupling between the tool and the OOMMF executable is relatively
loose (the OOMMF implementation internals can change without affecting
the Python interface) and transparent (developers and users can
inspect \texttt{mif} files for debugging purposes if required).
A disadvantage of this approach is that there are situations in which
CPU cycles are needed that could have been avoided through a more
tightly coupled interface that connects more directly to the OOMMF
internals.

\begin{figure}
  \includegraphics{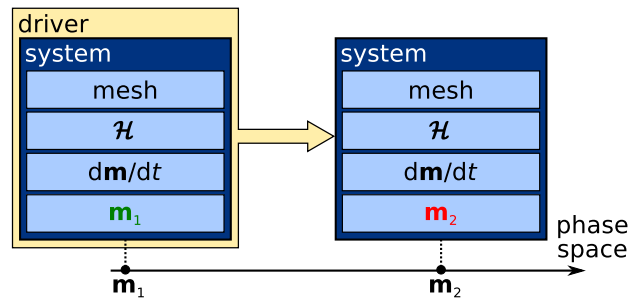}
  \caption{\label{fig:micromagneticmodel} A schematic representation
    of our micromagnetic model. The system object is completely
    defined with: (i)~mesh, (ii)~Hamiltonian $\mathcal{H}$, (iii)
    dynamics equation $\text{d}\mathbf{m}/\text{d}t$, and (iv)
    magnetisation configuration $\mathbf{m}$. The driver is
    ``driving'' (moving) the
    system through phase space by changing its magnetisation from
    $\mathbf{m}_{1}$ to $ \mathbf{m}_{2}$.}
\end{figure}

\paragraph*{Micromagnetic model description language.}
Here, we summarise the design of the micromagnetic model language that
is exposed to the user at the Python level.  There is not sufficient
space to detail our design and reasoning in this article. The full
code and growing documentation is available.~\cite{oommfc}
In summary, we base the micromagnetic model around a ``system'' which
is defined by providing: (i) mesh, (ii) Hamiltonian $\mathcal{H}$,
(iii) optional dynamics equation $\text{d}\mathbf{m}/\text{d}t$, and
(iv) the current magnetisation configuration $\mathbf{m}$. The mesh
contains the geometry and discretisation information, the Hamiltonian
captures the relevant interactions contributing to the energy, and for
time dependent problems there is an equation of motion.
To change the magnetisation, we use ``drivers'' (following terminology
introduced by the OOMMF team~\cite{Donahue1999}). Drivers
``drive'' the system in phase space by changing its
magnetisation. This can be either energy minimisation (e.g. conjugate
gradient) or time evolution through integration of the
Landau-Lifshitz-Gilbert (LLG)\cite{Gilbert2004} equation. We show
the schematic representation of our micromagnetic model in
Fig.~\ref{fig:micromagneticmodel}.
Based on this micromagnetic model, we have implemented the ``OOMMF
Calculator'' (\texttt{oommfc}) which can carry out the required
micromagnetic computations. When required,
the OOMMF calculator writes a mif file, calls OOMMF to execute it,
extracts the required information from the output files, and makes the
results available within the Python environment.

\section{Example}
\label{sec:example}
To illustrate the use of the Python interface for OOMMF, we
solve Standard problem~\cite{McMichael} 4 by computing the
magnetisation evolution in the thin film with $3 \,\text{nm}$
thickness of length $L = 500 \,\text{nm}$ and width
$d = 125 \,\text{nm}$. The material is Permalloy with magnetisation
saturation $M_\text{s}$ and exchange energy constant $A$.

As the first step, we need to import the Python module \texttt{oommfc}
(the name standing for OOMMF Calculator) which provides the commands
for defining the micromagnetic system and running OOMMF simulations,
as well as the \texttt{discretisedfield} module that we use for
defining the finite difference mesh and fields.
\begin{lstlisting}
import oommfc as oc
import discretisedfield as df
\end{lstlisting}

In our micromagnetic model, a system is defined completely by
providing its mesh, Hamiltonian, dynamics equation, and magnetisation
configuration. We specify the geometry through coordinates of two
points between which the cuboidal domain spans
and the size of a discretisation cell
($d_{x}, d_{y}, d_{z} = 2.5 \,\text{nm}, 2.5 \,\text{nm}, 3 \,\text{nm}$).
\begin{lstlisting}
L, d, th = 500e-9, 125e-9, 3e-9   # (m)
cellsize = (2.5e-9, 2.5e-9, 3e-9)  # (m)
mesh = oc.Mesh((0, 0, 0), (L, d, th), cellsize)
\end{lstlisting}
The variable \texttt{name} in the system object labels the
directory structure that holds the OOMMF output files.
\begin{lstlisting}
system = oc.System(name="stdprob4")
\end{lstlisting}

According to the standard problem 4 specification, the system's
Hamiltonian contains ferromagnetic exchange and demagnetisation energy
terms
\begin{equation}
  \mathcal{H} = \underbrace{A[(\nabla m_{x})^{2} + (\nabla m_{y})^{2} + (\nabla m_{z})^{2}]}_\texttt{Exchange(A)} + \underbrace{w_\text{d}}_\texttt{Demag()},
\end{equation}
where $m_{x}$, $m_{y}$, and $m_{z}$ are Cartesian coordinates of unit
magnetisation vector $\mathbf{m} =
\mathbf{M}/M_\text{s}$. We provide this Hamiltonian to the system object.
\begin{lstlisting}
A = 1.3e-11  # (J/m)
system.hamiltonian = oc.Exchange(A) + oc.Demag()
\end{lstlisting}

Next, we define the dynamics of the system, which is  governed by the
Landau-Lifshitz and Gilbert~\cite{Gilbert2004} equation which consists of
two (precession and damping) terms
\begin{equation}
  \frac{\text{d}\mathbf{m}}{\text{d}t} = \underbrace{-\gamma_{0}^{*}(\mathbf{m} \times \mathbf{H}_\text{eff})}_\texttt{Precession(gamma)} + \underbrace{\alpha(\mathbf{m} \times \frac{\text{d}\mathbf{m}}{\text{d}t})}_\texttt{Damping(alpha)},
\end{equation}
where $\mathbf{H}_\text{eff}$ is the effective field computed from the
system's Hamiltonian. We specify the equation of motion:
\begin{lstlisting}
gamma = 2.211e5  # (m/As)
alpha = 0.02
system.dynamics = oc.Precession(gamma) + \
                    oc.Damping(alpha)
\end{lstlisting}

In order to complete the definition of our micromagnetic model, we
 specify the initial magnetisation configuration $\mathbf{m}$,
 which is uniform in direction $(1, 0.25, 0.1)$, normalised to the
 value of $M_\text{s}$.
\begin{lstlisting}
Ms = 8e5  # (A/m)
system.m = df.Field(mesh, value=(1, 0.25, 0.1),
                      normalisedto=Ms)
\end{lstlisting}

In the first stage of standard problem 4, we need to relax the
system, and we use an energy minimisation driver. First we create the
\texttt{MinDriver} object and then pass the system object to the
\texttt{drive} method.
\begin{lstlisting}
md = oc.MinDriver()
md.drive(system)  # updates system.m in-place
\end{lstlisting}

After having called the minimisation driver, the system object now
contains the (equilibrium) magnetisation configuration for which the
system's energy is minimised.

So far, our micromagnetic system was at zero external magnetic
field. In the second stage we need to add the external magnetic field
$\mathbf{B} = (-24.6, 4.3, 0.0) \,\text{mT}$. Therefore, we add the
Zeeman energy density term
$w_\text{z} = -\mu_{0}\mathbf{M} \cdot \mathbf{H}$ to the
system's Hamiltonian, where $\mathbf{H} = \mathbf{B}/\mu_{0}$, with
$\mu_{0}$ being the magnetic constant. We carry on using the same
micromagnetic system as in the
relaxation step, but modify it by adding the Zeeman term to its
Hamiltonian. Nothing else changes; in particular the magnetisation
$\mathbf{m}$ is the same as after the relaxation stage.
\begin{lstlisting}
H = (-24.6e-3/oc.mu0, 4.3e-3/oc.mu0, 0)
system.hamiltonian += oc.Zeeman(H)
\end{lstlisting}

Now, we drive the system using a \texttt{TimeDriver} for
$1 \,\text{ns}$ and instruct the system to remember its magnetisation
evolution at 200 points during the nanosecond.
\begin{lstlisting}
td = oc.TimeDriver()
td.drive(system, t=1e-9, n=200)
\end{lstlisting}

\begin{figure}
  \includegraphics{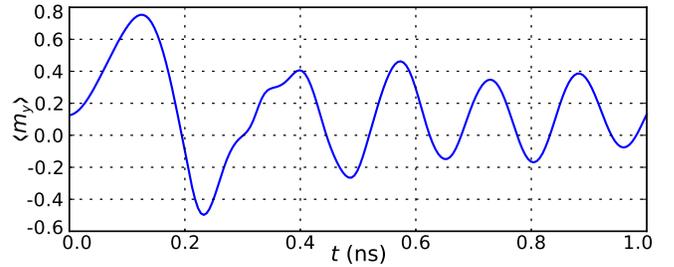}
  \caption{\label{fig:stdprob4} The average $y$ component of magnetisation
    time evolution for the first $1 \,\text{ns}$ computed with OOMMF
    simulation tool via our Python interface.}
\end{figure}

Finally, we can plot and save the time evolution of the average $y$ component
of magnetisation as shown in Fig.~\ref{fig:stdprob4}.
\begin{lstlisting}
myplot = system.dt.plot("t", "my")
myplot.figure.savefig("stdprob4-t-my.pdf")
\end{lstlisting}
The data analysis can be decoupled from running the simulation. In
that case, we parse the saved output files from OOMMF
runs and make the data available.

For multi-material simulations, we intend to use spatially varying
material parameters.

\section{Summary}
We summarise and discuss approaches to simulation user interface
design with computational micromagnetics as a case study. We argue
that embedding both the high level simulation commands and input
parameters as a domain specific language in an existing programming
language is the approach with most benefits. Consequently, we
implement such an interface that allows to drive OOMMF through the
Python programming language, and make it available as open
source.\cite{oommfc} We illustrated its use by solving the
micromagnetic standard problem~4.
We hope this interface can improve the micromagnetic simulation
workflows, supporting more reproducible and effective computational
science.

The design of the Python interface for OOMMF is a prototype for a
generic (Python-based) specification language for micromagnetic
simulation problems, which in the future can be extended to use other
computational backends in addition to OOMMF.

\begin{acknowledgments}
  This work was financially supported by the OpenDreamKit -- Horizon
  2020 European Research Infrastructure project (676541), the EPSRC's
  Centre for Doctoral Training in Next Generation Computational
  Modelling grant EP/L015382/1, and the EPSRC's Programme grant on
  Skyrmionics (EP/N032128/1).

  We thank Michael Donahue, Donald Porter, and Dmitri Chernyshenko for
  helpful discussions.
\end{acknowledgments}

%

\end{document}